\documentclass[%
 reprint,
 amsmath,amssymb,
 aps,
]{revtex4-1}

\usepackage{graphicx}
\usepackage{dcolumn}
\usepackage{bm}
\usepackage{epstopdf}
\usepackage{subfigure}

\begin{document}

\preprint{APS/123-QED}

\title{Gain-guided X-ray free-electron laser oscillator}

\author{Kai Li$^{1,2}$}
\author{Haixiao Deng $^{1,}$}%
 \email{denghaixiao@sinap.ac.cn}
 \affiliation{%
 $^1$Shanghai Institute of Applied Physics, Chinese Academy of Sciences, Shanghai 201800, China\\
 $^2$University of Chinese Academy of Sciences, Beijing 100049, China.
}%

\date{\today}

\begin{abstract}
The X-ray free-electron laser oscillator (XFELO) has recently been reconsidered a promising candidate for producing high-brightness, fully coherent pulses in the hard X-ray regime. This letter proposes a gain-guided XFELO scheme, avoiding the external focusing elements required for conventional configuration. Self-consistent theoretical analysis and three-dimensional numerical simulation results verify that the X-ray transverse mode in such an ``unstable" cavity is stable and robust due to electron beam gain-guiding. This scheme is capable of generating 14.3 keV photons of FEL radiation with peak power on the order of a gigawatt using Shanghai coherent light source parameters. The output pulse energy and transverse and longitudinal coherence are comparable with those from conventional XFELOs. This promising scheme is expected to contribute significantly to the construction and operation of a real XFELO.
\begin{description}

\item[PACS numbers]
41.60.Cr
\end{description}
\end{abstract}

\pacs{Valid PACS appear here}
\maketitle

X-ray free-electron laser (FEL) facilities around the world (see \cite{emma2010first,ishikawa2012compact,kang2017hard,altarelli2011european,milne2017swissfel}) generating X-ray pulses with peak brightness approximately 10 orders of magnitude higher than synchrotron radiation have enabled a broad scope of scientific investigations in fields such as biology; chemistry; material science; and atomic, molecular, and optical physics \cite{bostedt2016linac}. All cited hard X-ray FELs use self-amplified spontaneous emission (SASE) \cite{bonifacio1984collective} as their lasing mode. SASE starts with the initial electron beam shot noise and results in radiation with excellent spatial but rather poor temporal coherence. The spectral brightness of SASE can be improved by using advanced self-seeding technology in the X-ray regime \cite{ratner2015experimental,amann2012demonstration}. However, large shot-to-shot fluctuations in self-seeding ($\sim$ 50\% r.m.s.) still limit the applications of X-ray FELs. Recently, the generation of high repetition rate ($\sim$ 1 MHz), ultra low emittance electron bunches from a superconducting accelerator \cite{weise2003superconducting} and the realization of high reflectivity Bragg crystals \cite{shvyd2003x,shvyd2010high} have paved an alternative path for generating fully coherent X-ray FEL pulses using low-gain oscillators \cite{kim2008proposal,dai2012proposal}. 

In oscillator FELs, optical pulses are trapped within an optical cavity where an undulator of appropriate length is inserted. FEL gains occur each time the optical pulse meets and travels through the undulator together with an electron bunch. This configuration is well studied theoretically and has been experimentally demonstrated in the terahertz and infrared regions by employing suitable reflectors for the cavity \cite{billardon1983first,van1993felix,thompson2012first}. For an X-ray FEL oscillator (XFELO), an X-ray cavity with sufficient round trip feedback might be constructed using a newly available high reflectivity Bragg crystal with low absorption, like sapphire or diamond. Meanwhile, FEL starts from noisy spontaneous radiation in XFELO are purified efficiently by crystal mirrors with high reflectivity in several meV bandwidths. The output characteristics of an XFELO, especially the enhanced longitudinal coherence and improved output stability, would thus be extraordinary and complementary to a SASE FEL. With a peak brilliance comparable to SASE and average brilliance three orders of magnitude higher than SASE,
the X-ray FEL oscillator opens new scientific opportunities in various research fields such as X-ray inelastic scattering spectroscopy and nuclear resonant scattering, bulk-sensitive Fermi surface studies, X-ray imaging with near atomic resolution, and X-ray photon correlation spectroscopy.

Although studies indicate the XFELO theory appears to be feasible, practical challenges such as crystal thermal loading effects and X-ray optics remain \cite{kim2016oscillator,kolodziej2016diamond,song2016numerical}. For example, FEL oscillators require focusing components to control the transverse radiation profile and assure optimum gain. This typically involves the long wavelength regime, where curved mirrors are typically employed \cite{oepts1995free}. In the hard X-ray region, however, two available options are grazing-incidence curved mirrors or Be compound refractive lens \cite{lindberg2011performance,yumoto2013focusing,lengeler1999imaging}. These additional X-ray elements increase machine costs, facility complexity, and may even damage XFELO operational stability. In this letter, an XFELO scheme capable of functioning without external focusing elements is proposed, taking advantage of the large Rayleigh length of the X-ray and FEL gain-guiding effects.  
 \begin{figure}
  \centering
  \includegraphics[width=9cm]{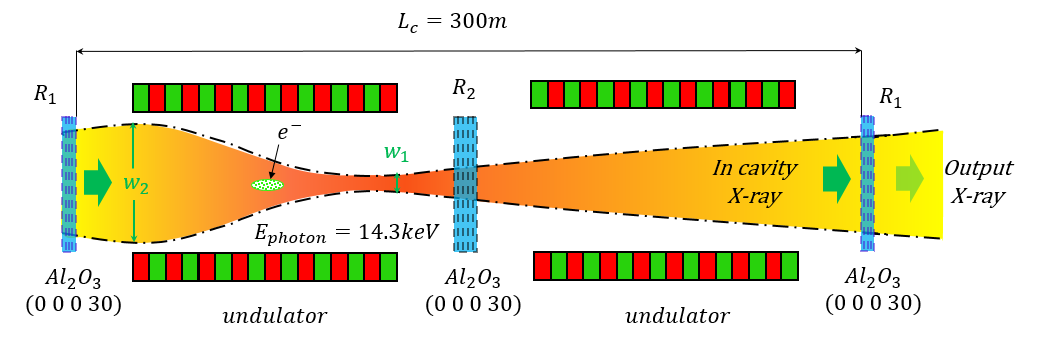}\\
  \caption{Scheme for gain-guided X-ray FEL oscillators. Dashed black line represents transverse envelop of X-ray radiation (shrinks inside the undulator due to FEL gain, expands during propagation in the vacuum). The overall system is displayed periodically along the X-ray pulse propagation. The left portion shows the X-ray forward trip with FEL gain, whereas the right displays the backward trip of the reflected X-ray.}\label{scheme}
\end{figure}

The basic schematic diagram is shown in Fig.~\ref{scheme}. For illustration, two parallel Bragg crystals are employed to form a symmetry X-ray resonator. The scheme might be implemented directly or with a more complex four-mirror cavity, enhancing wavelength tunability \cite{kim2009tunable}. The crystal thickness is adjusted to efficiently couple out X-ray power from the downstream mirror as well as maintain sufficient round trip reflectivity. The high repetition rate electron bunches match with the circulating X-ray pulse and supply sufficient gain to overcome the round-trip loss. Spontaneous radiation from leading electron bunches inside the undulators starts the XFELO, and a portion of this signal is reflected back and amplified by interactions with a later fresh electron beam. The entire system functions at a low gain regime where the single pass gain remains constant and the intra-cavity X-ray power experiences exponential growth. In the saturation regime, however, the strong electromagnetic field results in electron beam over-modulation and FEL gain degradation. When the energy coupled outside cavity is exactly compensated by the FEL single-pass gain (i.e. the zero net round trip gain), XFELO reaches equilibrium and delivers constant output power. As mentioned above, the X-ray spectrum is filtered during Bragg reflection and thus the longitudinal profile is mainly determined by the electron beam profile and crystal Darwin bandwidth. As this proposal lacks external focusing elements, the transverse mode is determined by the competition between the FEL gain-guiding effect (shrinking the X-ray profile) and X-ray diffraction during propagation. The orange region in Fig.~\ref{scheme} represents the X-ray transverse profile. Only the central portion of the field overlapping the electron beam receives the FEL gain $G_c$ inside the undulators. This transverse partial FEL gain leads to a single pass gain $G$ across the full X-ray profile. Outside the undulator segments, the X-ray profile expands and fills the peripheral region. To investigate this interesting phenomena, a theoretical model ignoring the influence of crystal diffraction is established to analyze transverse mode evolution \cite{colson1983multimode,pinhasi1996theory}. Assuming a Gaussian transverse profile electron beam with standard deviation of $\sigma_e$, the time and space variation of X-ray electric fields $E$ can be written as:
\begin{equation}
E(r;\,z)=\mathcal{R}e \left[ \tilde{u}(r;\,z)\, e^{j(\omega t-kz)} \right]
\label{eq1}
\end{equation}
where the wave vector is $k=2\pi/\lambda$ with a wavelength of $\lambda$ and angular frequency $\omega$, and the transverse complex phasor amplitude is $\tilde{u}$. At the exit of n-th pass through the undulator,
\begin{widetext}
\begin{equation}
\tilde{u}_n(r)=\sqrt{R\left( 1+G_c\,e^{-\frac{r^2}{2\sigma_e^2}} \right)}\,\frac{j\,k}{L_c}\int_0^\infty  \tilde{u}_{n-1}(r_0)\,e^{-j\frac{k(r^2+r_0^2)}{2L_c}}J_0(k\,r\,r_0/L_c)\,r_0\,dr_0
\label{eq2}
\end{equation}
\end{widetext}
where the round trip reflectivity is $R$. The system is assumed to have axial symmetry and Huygens' integral in cylindrical coordinates $r$ is used to depict the propagation of the paraxial wave through the round trip length $L_c$ \cite{siegman13lasers}. Ignoring phase modification, the X-ray pulse is then amplified by multiplying the FEL gain. At saturation, the gain-guided XFELO is expected to have stable output power, i.e. $\arrowvert\tilde{u}_{n}(r)\arrowvert^2=\arrowvert\tilde{u}_{n-1}(r)\arrowvert^2$. Since the X-ray signal is from an electron beam with a Gaussian profile, a Gaussian form solution is expected for the stable X-ray profile:
\begin{equation*}
\tilde{u}(r_;\,z)=u_0 \frac{e^{ j\,\psi(z)}}{w(z)}\, exp\left( -\frac{r^2}{w^2(z)}-jk\frac{r^2}{2R_r(z)} \right)
\label{eq3}
\end{equation*}
where $u_0$ is the electric field amplitude, $w(z)$ is the X-ray radius in which the amplitude falls to $1/e$, $R_r(z)$ is the radius of beam wavefront curvatures, and $\psi(z)$ is the Gouy phase. Inserting the Gaussian form solution into Eq.~(\ref{eq2}) and assuming $G_c\gg\,1$, the following is obtained: 
\begin{eqnarray}
\left( \frac{w_2}{w_1} \right)^2=R\,G_c,
\label{eq4}
\\
\frac{1}{4\sigma_e^2}+\frac{1}{w_2^2}=\frac{1}{w_1^2}.
\label{eq5}
\end{eqnarray}
where $w_2$ and $w_1$ are the X-ray radii at the entrance and exit of undulator, as shown in Fig.~\ref{scheme}. Instead of the traditional criteria $R\,G_c >1$ for FEL power growth, Eq.~(\ref{eq4}) implies the single pass gain at the beam center should compensate for the diffraction effect, whereas Eq.~(\ref{eq5}) shows that $w_1$ is smaller than $w_2$ (as expected) and reveals that X-ray profile shrinkage is indeed due to FEL gain. Note that since X-ray power is expected to be stable in the saturation regime, only the amplitude of $\tilde{u}$ was taken into account during the derivation of Eqs.~(\ref{eq4}) and (\ref{eq5}), and there may be no periodical solution for $\tilde{u}$ itself.

The XFELO extracts kinetic energy from electrons and converts it to X-ray pulses. The efficiency of the system can be estimated easily: according to \cite{kroll1981free,li2017gain}, the radiation peak power density is inversely proportional to the number of undulator periods $N_u$, making the XFELO efficiency:
\begin{equation}
\eta=K_1\left( \frac{w_1}{\sqrt{2}\sigma_e} \right)^2\frac{1}{N_u}
\label{eq6}
\end{equation}
where $K_1$ is a coefficient. According to Ref.~\cite{madey1971stimulated,colson1977one}, the FEL small signal gain is proportional to the square of the number of undulator periods multiplied by the electron density, which is inversely proportional to the transverse electron beam area for a given bunch charge, i.e.,
\begin{equation}
R\,G_c=K_2 \frac{N_u^2}{\sigma_e^2}
\label{eq7}
\end{equation}
where $K_2$ is the coefficient. Through Eqs.~(\ref{eq4}), (\ref{eq6}), and (\ref{eq7}), the efficiency ratio between gain-guided and conventional XFELO is seen to be:
\begin{equation}
\epsilon=\frac{\eta_{\rm gg}}{\eta_{\rm normal}}=\frac{w_1^3}{2w_2\sigma_e^2}
\label{eq8}
\end{equation}
where the radiation radius of the traditional XFELO with external focusing elements is assumed to equal the transverse electron bunch size for the maximizing coupling factor ($w_1=\sqrt{2}\sigma_e$). For simplicity, we further assume the X-ray waist is at the exit of the undulator, where the X-ray radius becomes minimum ($w_1=w_0$) and spot size parameter $w_2$ is given by:
\begin{equation}
w_2=w_0\sqrt{1+\left( \frac{L_c}{z_R} \right)^2}
\label{eq9}
\end{equation}
where the Rayleigh length is $z_R=\pi w_0^2/\lambda$.

\begin{figure}
  \centering
  \includegraphics[width=8cm]{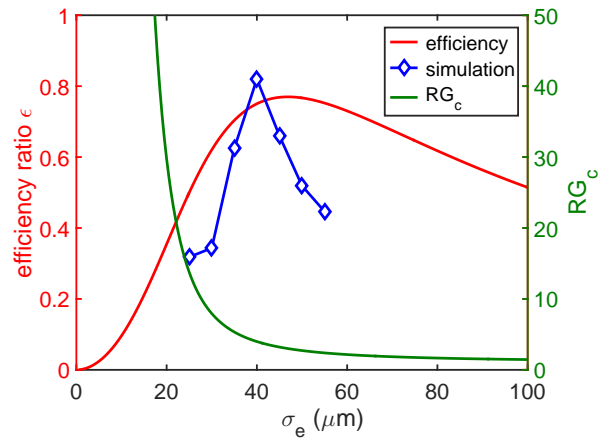}\\
  \caption{Theoretical efficiency ratio estimation of gain-guided to normal XFELO (red line), and corresponding simulation results (blue line with diamond marker). The minimum product $R\,G_c$ required for power growth is shown as a function of electron beam transverse size by the green line. }
\label{efficiency}
\end{figure}

Combing the previous equations, the ratio of efficiency can be solved as a function of electron beam transverse size, as shown in Fig.~\ref{efficiency}. In this letter, an XFELO at 14.3 keV photon with a round trip length $L_c=300$ m resonator is considered. Efficiency increases as the electron beam size increases, reaching a maximum value of 77\% at $\sigma_e=43 \mu$ m. Low charge density decreases efficiency at large electron beam sizes. Additionally, $w_2/w_1$ decreases as the transverse electron beam size increases, as did the required product $R\,G_c$. The relatively larger required FEL gain $G_c$ might resemble RAFEL \cite{huang2006fully}. However, the round trip net gain $G$ remained at the level of a low gain oscillator due to X-ray diffraction. This letter concentrates on the X-ray transverse profile evolution and explores the possibility of a self-focusing XFELO scheme instead of trying to focus X-rays with external elements. The proposed scheme provides a moderate gain at the transverse X-ray center and controls the transverse mode of the X-ray via cavity configuration, thus benefiting from the high stability of a small XFELO gain.

As a numerical example, the proposed gain-guided XFELO is simulated using the parameters of the Shanghai Coherent Light Facility (SCLF), the first hard X-ray FEL facility in China (currently under construction). The 1 MHz repetition rate, 8 GeV electron bunches with ultra-low 0.4 $\mu$m-rad normalized emittances delivered by the CW superconducting accelerator are suitable for XFELO operation. The bunch charge is 100 pC and peak current is 1 kA. As mentioned above, X-ray photon energy is set to 14.3 keV, equal to the Bragg energy of sapphire crystal mirrors at normal incidence to the (0 0 0 30) atomic planes. For a 0.01\% slice relative energy spread electron beam, seven segments of the 5 m undulators with 26 mm period are used to provide sufficient FEL gain. A typical FODO lattice is employed to control the electron beam size. The three-dimensional simulation is conducted using a combination of GENESIS \cite{reiche1999genesis}, OPC \cite{van2009time}, and BRIGHT \cite{kai2017systematical}. The simulation results of various electron beam sizes with carefully optimized mirror reflectivity for power maximization are shown in Fig.~\ref{efficiency}. As expected, the peak power efficiency first increases and then decreases as the transverse beam size increases, reaching its maximum $\epsilon=82\%$ when $\sigma_e=40 \mu m$, agreeing well with the theoretical study. The faster simulated decline following optimum electron beam size might be due to the three-dimensional effects leading to deviations in FEL gain calculations in the model.

\begin{figure*}
   \centering
   \subfigure{\includegraphics*[width=160pt]{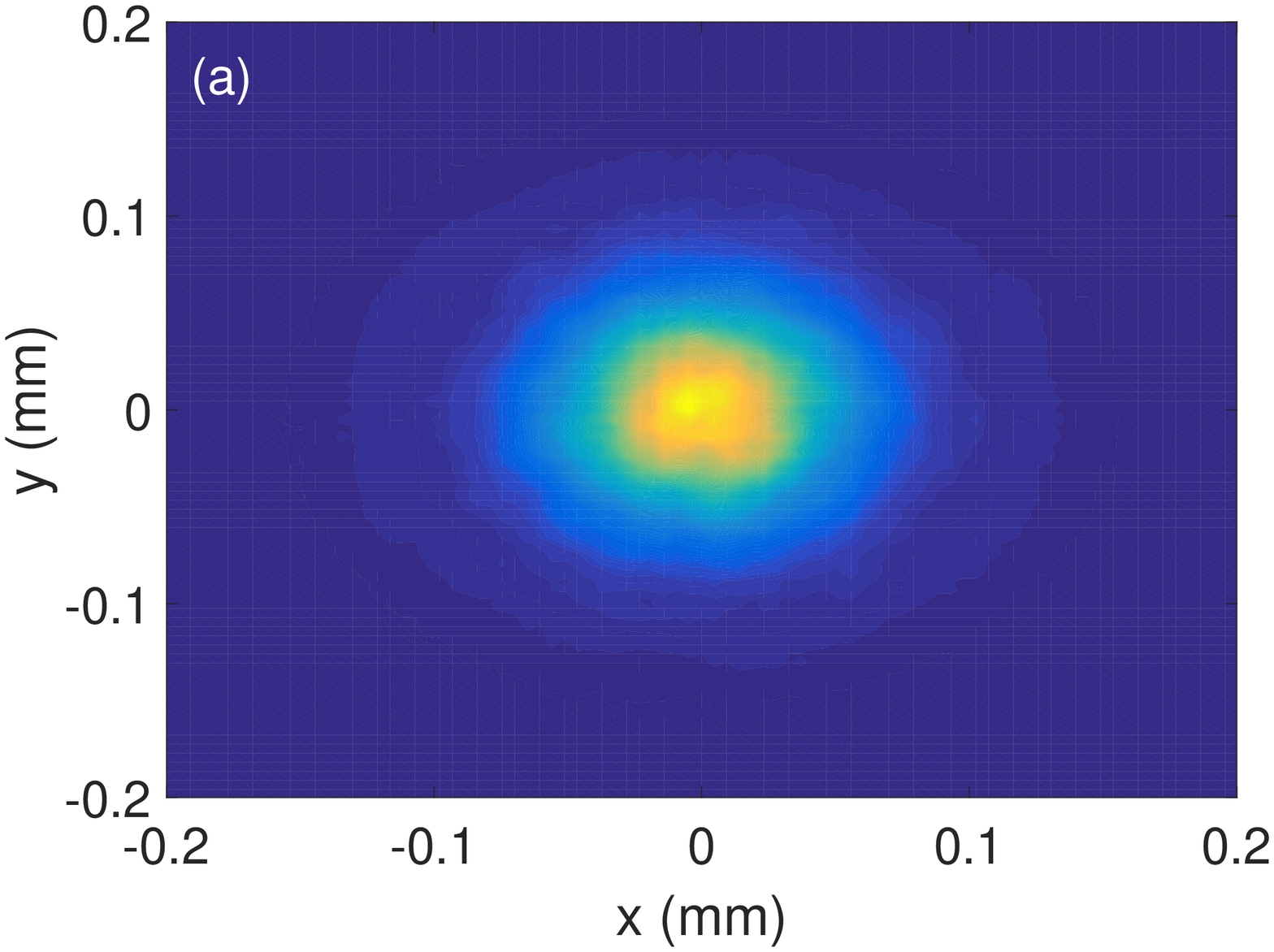}}
   \subfigure{\includegraphics*[width=160pt]{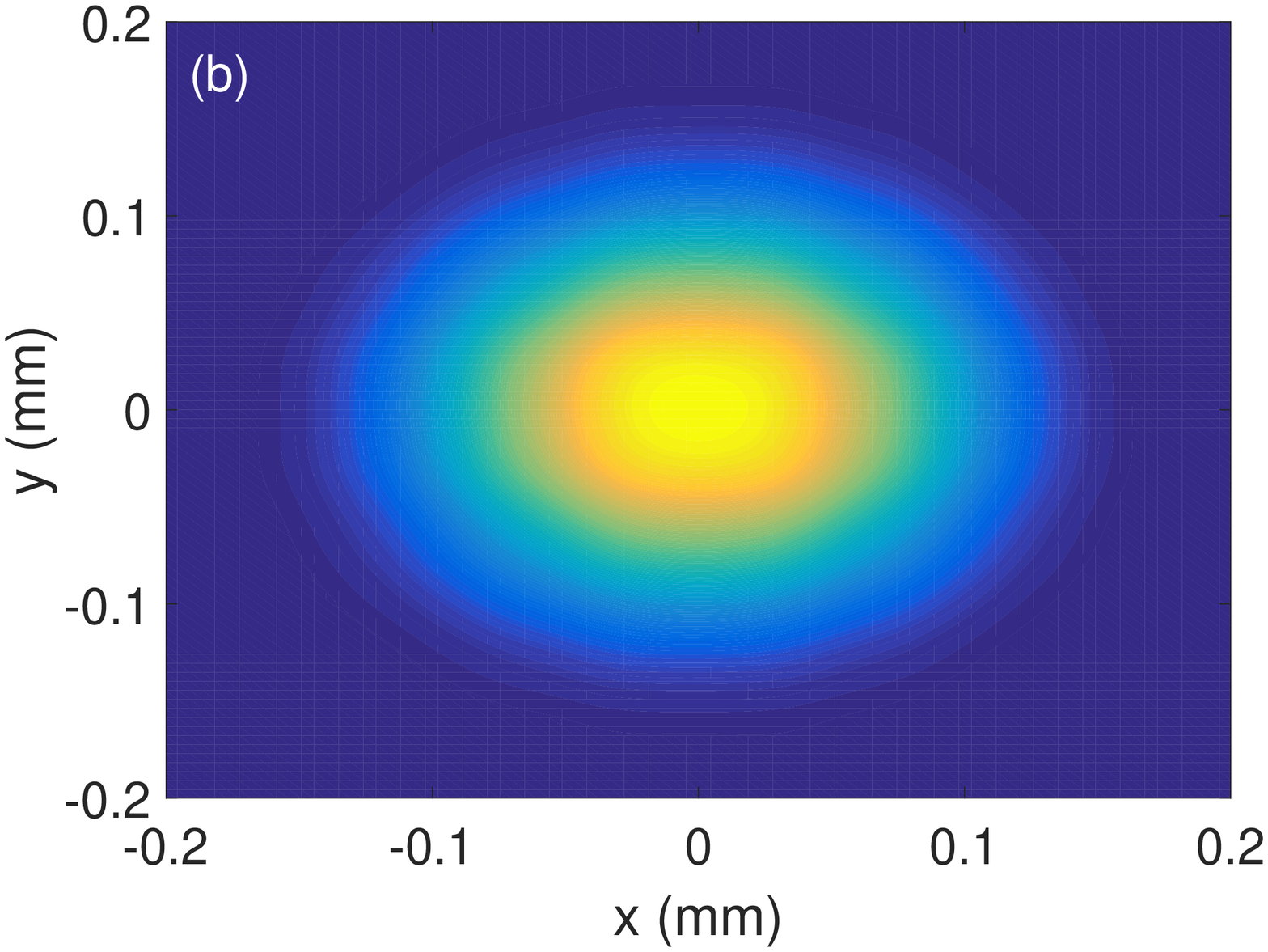}}
   \subfigure{\includegraphics*[width=160pt]{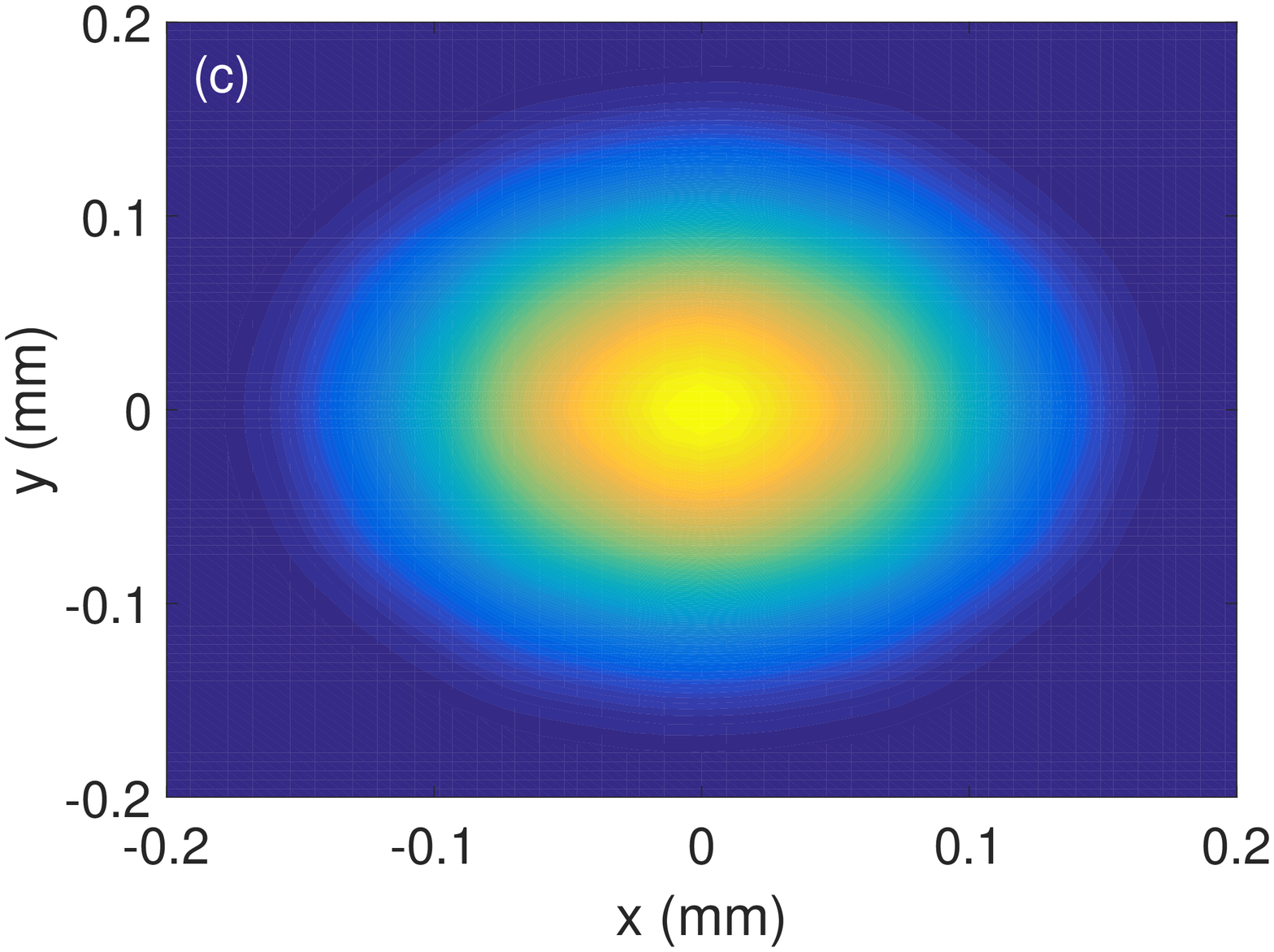}}
   \caption{Simulation results of X-ray transverse profile for gain-guided XFELO at different positions: (a) undulator exit; (b) upstream mirror; (c) entrance of undulator.}
   \label{trans}
\end{figure*}

\begin{figure*}
   \centering
   \subfigure{\includegraphics*[width=160pt]{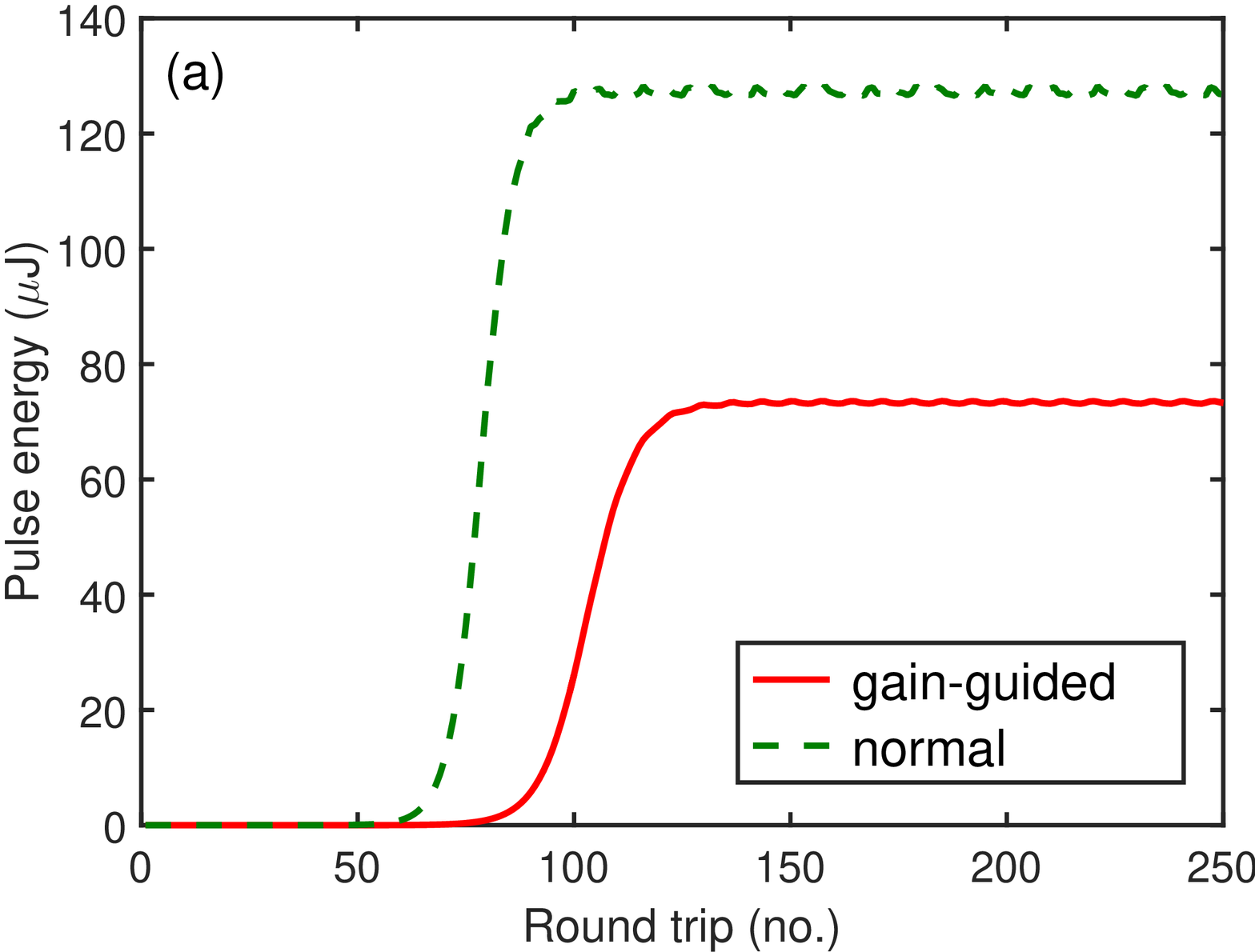}}
   \subfigure{\includegraphics*[width=160pt]{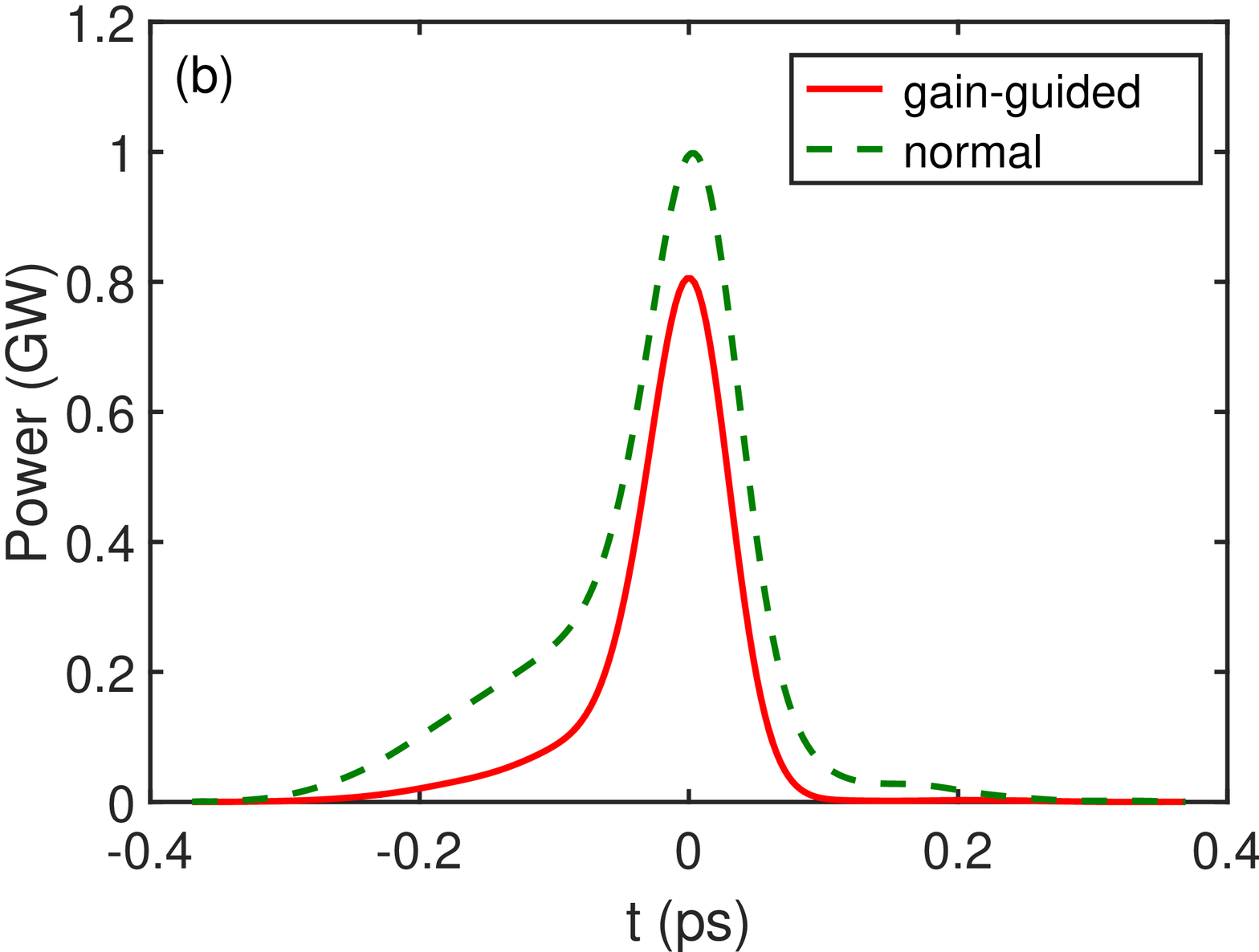}}
   \subfigure{\includegraphics*[width=160pt]{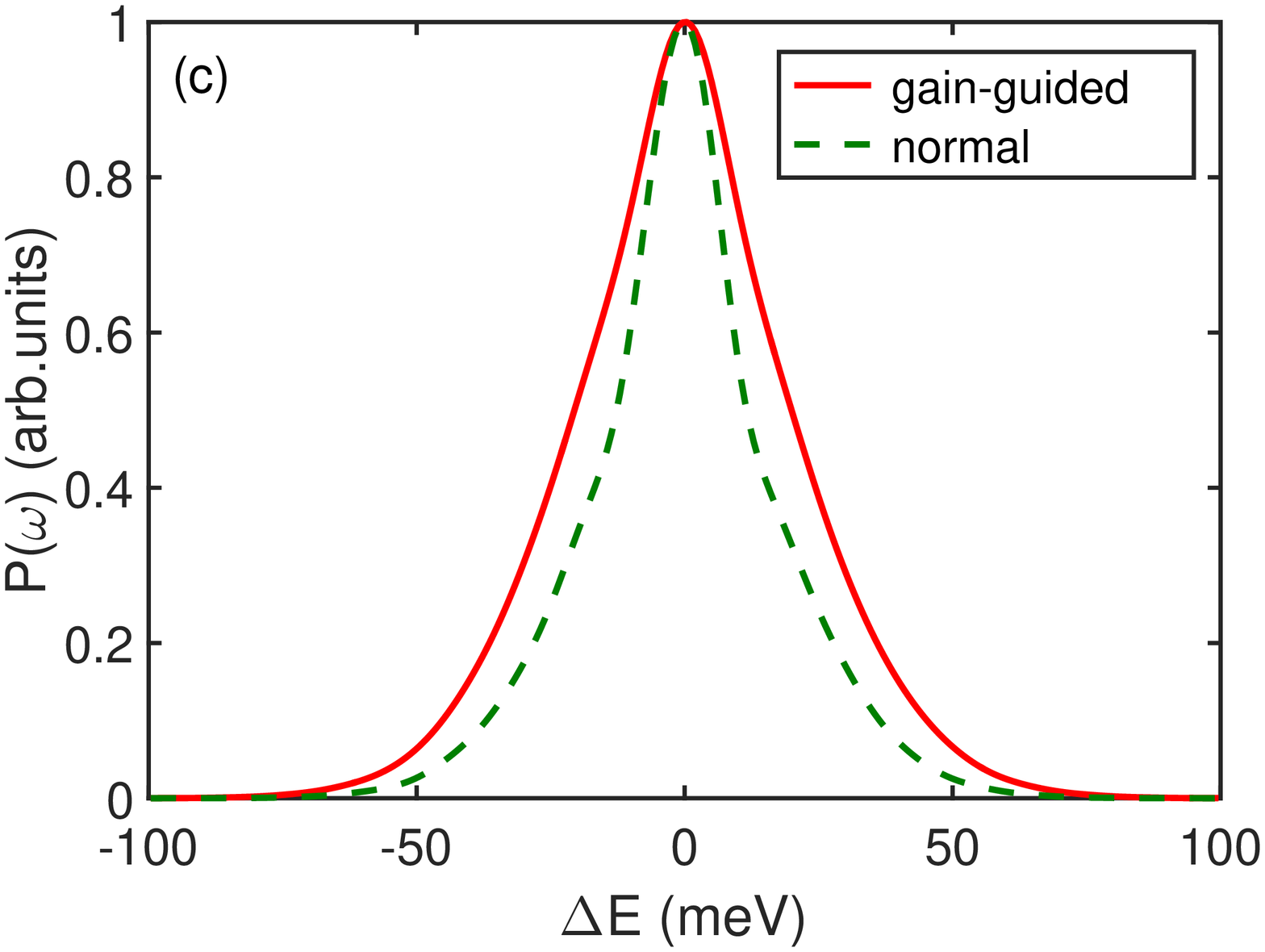}}
   \caption{Simulation results of normal XFELO and gain-guided XFELO: (a) X-ray pulse energy evolution at different round trip; (b) output temporal profile; (c) spectrum.}
   \label{results}
\end{figure*}

The simulation result details for maximum efficiency are shown in Figs.~\ref{trans} and \ref{results}. The downstream crystal thickness is $d_2=70 \mu$m, providing sufficient reflectivity ($R_2=95\%$), whereas the upstream mirror thickness is $d_1=40 \mu$m with a reflectivity of $R_1=70\%$ for efficient output coupling.  The additional loss is due to the cavity output coupling and Bragg crystal mirror bandwidth stop ($\sim 65\%$). The transverse radiation profile at the undulator exit, upstream mirror, and undulator entrance are shown in Fig.~\ref{trans} (a), (b) and (c), respectively. Since no focusing elements exist, the propagation of X-rays inside the vacuum chamber leads to a beam size expansion of nearly a factor of 2 during one round trip. The overall loss is exactly compensated by the FEL gain $G_c=28$ in the XFELO saturation regime. The parameter $M^2$ is calculated to evaluate the X-ray transverse quality from the gain-guided XFELO. An $M^2$ value close to unity indicates the X-ray pulse from the gain-guided XFELO could be perfectly focused onto a small spot in the experimental station. The output pulse energy at different round trips is presented in Fig.~\ref{results} (a), starts from the shot noise and growing exponentially before reaching a stable value of 73.6 $\mu$J. Figure~\ref{results} (b) and (c) show the output X-ray temporal profile and spectrum, respectively. The output peak power is nearly 0.8 GW, meaning the peak power efficiency ratio between gain-guided (red solid line) and normal XFELO (green dashed line) is approximately 82\%. The X-ray pulse duration is 75 fs (FWHM) with a bandwidth of 41 meV (FWHM), corresponding to a 0.75 time-bandwidth product (close to the value of the Fourier transform limit of a Gaussian pulse).

The robustness of gain-guided XFELOs is investigated in a further study. Due to the relatively large X-ray profile and partial overlap between electron beams and X-rays, the crystal tilt angular misalignment should be less than 50 nrad, much easier to achieve than the 10 nrad in a normal XFELO \cite{kim2009tunable}. This large X-ray profile also relaxes the requirements for electron beam trajectories along the undulators; for instance, the required 18 $\mu$m (r.m.s.) beam orbit straightening can be established by beam-based alignment. When compared to normal XFELOs, the X-ray footprint on the crystal surface is approximately doubled, which is helpful for reducing crystal thermal loading. Additionally, these discussions and simulations indicate that gain-guided XFELO would enable X-ray output power adjusting by simply tuning the quadruple magnet strength inserted between the undulators instead of the changing mirror reflectivity mechanically.

This letter proposed a gain-guided XFELO eliminating external focusing elements and thus simplifying X-ray cavity configurations. A theoretical model based on Huygens-Fresnel paraxial wave propagation is used to describe the transverse mode evolution of X-ray pulses. The electron beam plays two roles in X-ray optics: gain media and ``focusing element''. Taking advantage of the wave diffraction effect during propagation in the vacuum chamber, the X-ray power density at the center decreases and over-modulations of electron beams inside the undulator are mitigated, making the energy extraction efficiency of gain-guided XFELOs comparable to that of the conventional ones. With some reasonable assumptions, equations are obtained for predicting the requirements and expected performances of the proposed scheme. The feasibility of the proposal was verified through three-dimensional numerical simulations using SCLF parameters as an example. The typical results of the two-mirror symmetry cavity indicates that a gain-guided XFELO is capable of generating 73.6 $\mu$J, as large as 67.5\% pulse energy of a normal configuration (128 $\mu$J). The output peak power of 0.8 GW at 41 meV bandwidth are equivalent to that of conventional systems. Without inserting additional focusing elements, the proposed scheme holds several advantages over normal XFELO: avoidance of potential thermal loading effects, improved wavelength and peak power tuning, and enhanced system robustness. This proposal is expected to promote the future construction of XFELOs.

This work was partially supported by the National Natural Science Foundation of China (11775293), the National Key Research and Development Program of China (2016YFA0401900), the Young Elite Scientist Sponsorship Program by CAST (2015QNRC001) and Ten Thousand Talent Program.

\nocite{*}

\bibliography{references}

\end{document}